# CUPCF: Combining Users Preferences in Collaborative Filtering for Better Recommendation

**Mostafa Khalaji[1] . Nilufar Mohammadnejad[2]**

[1] *Faculty of Computer Engineering, K. N. Toosi University of Technology, Tehran, Iran*
*Tel: +98919-7337358, E-mail:* **Khalaji@email.kntu.ac.ir**

[2] *Islamic Azad University, Shahr-e-Qods Branch, Tehran, Iran*

**Abstract**

How to make the best decision between the opinions and tastes of your friends and acquaintances? Therefore, recommender systems are used to solve such issues. The common algorithms use a similarity measure to predict active users' tastes over a particular item. According to the cold start and data sparsity problems, these systems cannot predict and suggest particular items to users. In this paper, we introduce a new recommender system is able to find user preferences and based on it, provides the recommendations. Our proposed system called CUPCF is a combination of two similarity measures in collaborative filtering to solve the data sparsity problem and poor prediction (high prediction error rate) problems for better recommendation. The experimental results based on MovieLens dataset show that, combined with the preferences of the user's nearest neighbor, the proposed system error rate compared to a number of state-of-the-art recommendation methods improved. Furthermore, the results indicate the efficiency of CUPCF. The maximum improved error rate of the system is 15.5% and the maximum values of Accuracy, Precision and Recall of CUPCF are 0.91402, 0.91436 and 0.9974 respectively.

## 1. Introduction

Increasing instantaneous online information invites users to challenges. The confusion and waste of time users are obvious examples of these challenges. Fortunately, with the advancement of technology, this platform has been provided to track the behavior of users on the internet. Thus, recommender systems are able to deal with the mass of information users, to help them achieve their goals. By modeling users' behavior, discovering the tastes and preferences of them on webspace, one can no longer worry about the challenges that were posed. Nowadays, the importance of such systems and their use is evident.
Recommender systems are basic models that are based on them. One of the most versatile models is collaborative filtering. This model uses the users' rating to predict and offer products to active users. Accordingly, this model requires a user-item ratings matrix [1].
Each model, in turn, has its own problems that impact on the performance of the system. The main problems of this model include high error rate, cold start, data sparsity, and accuracy. The collaborative filtering uses the comments (ratings) and other preferences of users to suggest specific items to the active user. This model uses a similarity measure to find the nearest neighbor. Each of these similarity measures has the ability to solve the problems mentioned, one is able to partly solve the cold start problem and another one is able to improve system performance. Due to the sparse matrix, most of these models cannot provide suggestions for active users. In this paper, we have presented a recommender system based on collaborative filtering. Our proposed system called CUPCF is a combination of two similarity measures. CUPCF uses two similarity measures simultaneously as a new method for improving the error rate of the system. CUPCF does not just consider the similarity of neighbor users by a certain similarity measure. The proposed system was evaluated on MovieLens dataset and the results showed that the system error rate was improved compared to a number of state-of-the-art recommendation systems (other methods) that used one similarity measure for prediction and suggestions.

The structure of the paper is as follows. In the second part, we will review the workings of the researchers. The third part introduces the proposed systems. Fourth, the experimental results are compared with other methods in which the results will be discussed. Finally in the last part of the conclusions presented.

## 2. Related Works

Collaborative filtering is one of the most important and popular models in recommender systems. This model is divided into two main ways: model-based and memory-based [1]. The memory-based approach uses the method of finding the nearest neighbor to the active user and then predicts users' preferences and shows a list of suggestions in outbound. The performance of this model depends on the user-item ratings matrix [2]. The model-based approach makes a model of user behavior in the offline phase, and in the online phase, predicts and provides a list of recommendation. User-based and item-based collaborative filtering methods are used as two important techniques in the recommender systems [3]. The user-based technique offers items that have not yet been viewed by the active user (unrated) based on the nearest users [4]. On the other hand, the item-based technique predicts unrated items based on nearest neighbor items [5]. Therefore, similarity measures are used to find the nearest neighbor, each with advantages and disadvantages. Most of the traditional similarity measures include Pearson correlation coefficient [6], cosine measure [7] and mean squared difference [8]. Most of these similarity measures are not suitable for new users cold start problem.

Ahn proposed a new similarity measure called PIS (proximity, impact, and popularity) for solving the cold start problem [9]. Liu et al. proposed a new heuristic similarity measure (NHSM) to solve the cold start problem that could be improved collaborative filtering accuracy. Their method is effective when a small number of ratings are available. They modified the Jaccard [10] and the PIS [9] similarity measures [11]. Choi et al. introduced a new similarity measure to select neighbors for each item in the collaborative filtering. In their method, the rating of a user on an item is weighted by the item similarity between the item and the target item. [12]. Yang et al. proposed a collaborative filtering model based on heuristic formulated inferences. They introduced a new similarity measure and considered users' preferences and rating patterns for improving the prediction quality [13]. Zhang et al. proposed an effective way to improve the ability of finding the nearest neighbor and reliable for each active users. Their aim was to provide an effective model-based recommender system to solve the data sparsity problem [14]. In 2015, Park et.al proposed a fast collaborative filtering model using nearest neighbor graph which called RCF to reduce time complexity. Their model reversed the process of finding the nearest neighbor in collaborative filtering [3]. Bellogin introduced methods to improve the performance of the recommender systems, which selected Herlocker's weighting (HW) and McLaughlin's weighting (MW) methods to determine which users were closely related to the user's tastes [15]. Fangyi Hu in 2018 introduced a three-segment similarity measure method for collaborative filtering model. She/He improved the performance of similarity measure by computing the similarity between users based on the number of user ratings along with item similarity and user attribute similarity [16].

Nadi et al. proposed a recommender system based on fuzzy ant colony (FARS). They combined collaborative and content-based filtering for better recommendation [17]. Javari et al. proposed a recommender system based on collaborative filtering and resources allocation for improving the performance of their system. Using the resource allocation method, they were able to obtain the degree of confidence of each user based on the similarity achieved [18]. In 2017, Khalaji designed a hybrid recommender system based on neural network and resource allocation that solved the scalability and cold start problems [19]. In 2019, Khalaji et al. introduced a recommender system based on fuzzy clustering and heuristic similarity measure called FNHSM_HRS for improving the system performance, especially for scalability and cold start problems. Their method had two phases: offline and online. In first, the users clustered by a fuzzy c-means method, then the unobserved items were predicted by NHSM similarity measure [20]. In 2019, Khalaji and Mohammadnejad. introduced a hybrid movie recommender system called FCNHSMRA_HRS, a combination of fuzzy clustering, heuristic similarity measure, and resource allocation methods to solve the scalability and cold start problems and improved the performance of their system [21]. In 2019, Khodaverdi et al. proposed a movie hybrid recommender system based on clustering and popularity. Their system clustered the users who were similar to each other by using the K-means clustering method and using ratings popularity to predict the users' preferences to specific movies [22]. In 2019, Khalaji proposed a new recommender system called NWS_RS for movie recommendation. His method was able to personalize the recommendation by segmenting users' age. NWS_RS used the new weighted similarity (NWS) for improving the accuracy of prediction of unobserved movies for active users. NWS_RS managed the scalability problem and solved the data sparsity problem [23].

## 3. The proposed system CUPCF

Fig. 1 shows the structure of the proposed CUPCF system. This structure is based on user-based collaborative filtering model.

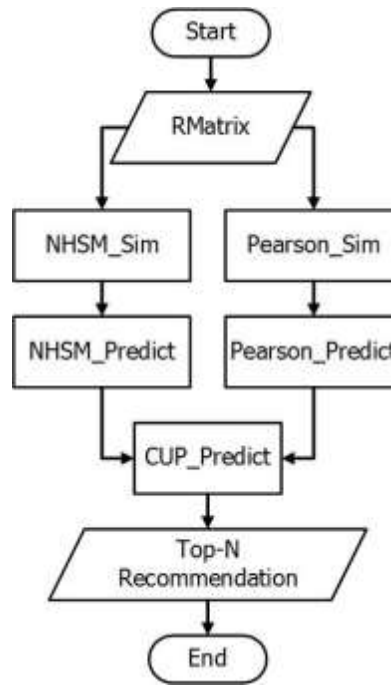

**Fig. 1** The CUPCF structure.

Recommender systems have a user-item rating matrix which includes a number of users' ratings for items. In this paper, we demonstrate users with $U = [u_1, u_2, ..., u_m]$, items with $I = [i_1, i_2, ..., i_n]$ and the matrix of ratings named RMatrix. The RMatrix size equals the number of users × the number of items that is $N \times M$.

Memory-based systems have two phases, prediction and recommendation. These systems use the nearest neighbors' preferences to predict unrated items for active users. Therefore, it is required to find the nearest users as neighbors with active users. To find these users, the different similarity measures are used. For example, the basic model of memory-based systems is depicted in Fig. 2.

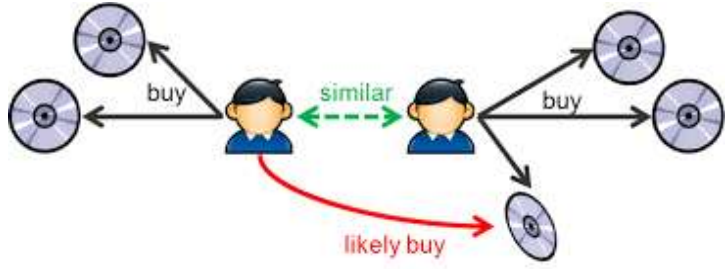

**Fig. 2** The basic model of memory-based systems.

3.1 Find K-Nearest Neighbors

Data sparsity problem arises from the phenomenon that users in general rate only a limited number of items [24]. This leads to very low accurate similarity based on the previous similarity measures. Therefore, NHSM [11] is a new user similarity measure to improve the recommendation and prediction performance when only few ratings are available for calculating the similarity for all users. This similarity measure not only considers the local context information of user ratings, but also the global preference of user behavior. The NHSM can overcome the drawback of the Pearson similarity measure when the rating matrix is sparse.

CUPCF receives RMatrix as an input, then uses the Pearson and NHSM similarity measures to determine k-neighbors of each user. Similarities between active users and other users based on NHSM is calculated according to Eq. (1) and this measure has two main coefficients.

$$NHSM\_Sim(u,v) = JPSS_{Sim(u,v)} \cdot URP_{Sim(u,v)} \tag{1}$$

To calculate the $NHSM\_Sim$ similarity measure, the $JPSS_{Sim(u,v)}$ similarity measure should first be calculated, which itself is derived from two similarity measures that are mentioned in Eq. (2) and Eq. (3).

$$JPSS_{Sim(u,v)} = PSS_{Sim(u,v)} \cdot Jaccard'_{Sim(u,v)} \tag{2}$$

$Jaccard'_{Sim(u,v)}$ similarity is an improved formula from traditional Jaccard similarity measure. For considering the proportion of the common ratings and improving the accuracy (high accuracy), the $Jaccard'_{Sim(u,v)}$ is used.

$$Jaccard'_{Sim(u,v)} = \frac{|I_u \cap I_v|}{|I_u| \times |I_v|} \tag{3}$$

$|I_u \cap I_v|$ represents the number of similar items that users $u$ and $v$ have seen and $|I_u|$ indicates the number of items that the active user ($u$) and $|I_v|$ indicates the number of items the other user ($v$) have rated. The $PSS_{Sim(u,v)}$ measure is composed of three factors of similarity, and obtained through Eq. (4).

$$PSS_{Sim}(r_{u,p}, r_{v,p}) = Proximity(r_{u,p}, r_{v,p}) \cdot Significance(r_{u,p}, r_{v,p}) \cdot Singularity(r_{u,p}, r_{v,p}) \tag{4}$$

The Proximity factor calculates the absolute difference between two ratings. The second factor is Significance. We assume that the ratings are more significance if two ratings are more distant from the median rating. For example, if two user rate two items as (4, 4) or (2, 2). We think it is more significant than two user give (5, 3) or (4, 2). The third factor is called the Singularity. This factor represents how two ratings are different from other ratings The $Proximity$, $Significance$, and $Singularity$ are calculated through Eq. (5-7).

$$Proximity(r_{u,p}, r_{v,p}) = 1 - \frac{1}{1 + \exp(-|r_{u,p} - r_{v,p}|)} \tag{5}$$

$$Significance(r_{u,p}, r_{v,p}) = \frac{1}{1 + \exp(-|r_{u,p} - r_{med}| \cdot |r_{v,p} - r_{med}|)} \quad (6)$$

$$Singularity(r_{u,p}, r_{v,p}) = 1 - \frac{1}{1 + \exp(-|\frac{r_{u,p} + r_{v,p}}{2} - \mu_p|)} \quad (7)$$

Where $r_{u,p}$ is the rating of item $p$ by user $u$ and $r_{v,p}$ is the rating of item $p$ by user $v$. $r_{med}$ is the median value in the rating scale. The rating matrix in the CUPCF recommender system has a scale of 1-5, with an average of 3. $\mu_p$ is also the average rating of item $p$ by users. The last step in Eq. (1) is to calculate the similarity measure of $URP_{Sim(u,v)}$, which is obtained by using Eq. (8). This measure is the preference of each user. Different users have different rating preferences. Some users prefer to give high ratings. On the other hand, some users tend to rate low value. In order to reflect this behavior preference, NHSM uses the mean and variance of the rating to model the user preference.

$$URP_{Sim(r_{u,p}, r_{v,p})} = 1 - \frac{1}{1 + \exp(-|\mu_u - \mu_v| \cdot |\sigma_u - \sigma_v|)} \quad (8)$$

Where $\mu_u$ and $\mu_v$ is the mean of rating of user $u$ and $v$ respectively. $\sigma_u$ and $\sigma_v$ represent the standard variance of user $u$ and $v$, which is obtained in accordance with Eq. (9).

$$\sigma_u = \sqrt{\sum_{p \in I_u} \frac{(r_{u,p} - \bar{r}_u)^2}{|I_u|}} \quad (9)$$

The value of $NHSM\_Sim(u, v)$ ranges between 0 to 1. The value 1 indicates the most similarity between users and 0 shows the difference between two users.
After that, the similarity between the active user and other users based Pearson correlation coefficient is calculated according to Eq. (10).

$$Pearson\_Sim(u, v) = \frac{\sum_{p \in I_u \cap I_v}(r_{u,p} - \mu_u) \cdot (r_{v,p} - \mu_v)}{\sqrt{\sum_{p \in I_u \cap I_v}(r_{u,p} - \mu_u)^2} \cdot \sqrt{\sum_{p \in I_u \cap I_v}(r_{v,p} - \mu_v)^2}} \quad (10)$$

Where $r_{u,p}$ represents the rating of active user $u$ for item $p$ and $\mu_u$ is the average rating given by user $u$. The value of $Pearson\_Sim(u, v)$ ranges between -1 to 1. The value 1 indicates the most similarity between users and -1 shows the difference between two users. The similarity calculated affected by the number of shared rated items between users.
After calculating the similarity between users, based on the two different measures as mentioned, two separate matrices called $NHSM\_Sim$ and $Pearson\_Sim$ are generated at the M × N number. These matrices are symmetric and the elements of matrices demonstrate the degree of users' similarity with active users. For example, if $u1$ is an active user and $u2$ is a neighbor user, the similarity between $u1$ and $u2$ is 0.0122, given the symmetry of the matrix, the similarity between user $u2$ and active user $u1$ is 0.0122. Table 1 is a simple example of RMatrix, and Table 2 and Table 3 show the NHSM and Pearson similarity functions.

**Table 1** An example of the user-item rating matrix (RMatrix). The missing ratings are represented by the symbol ?.

|    | i1 | i2 | i3 | i4 |
|----|----|----|----|----|
| **u1** | ? | 5 | 4 | 3 |
| **u2** | 4 | 4 | ? | 2 |
| **u3** | ? | ? | 1 | 4 |
| **u4** | 5 | 2 | 4 | 4 |
| **u5** | 1 | ? | 3 | ? |

**Table 2** An example of the users similarity matrix, according to NHSM similarity measure

| NHSM Similarity Measure | | | | |
|----|----|----|----|----|
|    | u2 | u3 | u4 | u5 |
| **u1** | 0.0122 | 0.0066 | 0.0213 | 0.0035 |
| **u2** |  | 0.0025 | 0.0150 | 0.0010 |
| **u3** |  |  | 0.0125 | 0.0018 |
| **u4** |  |  |  | 0.0067 |

**Table 3** An example of the users similarity matrix, according to Pearson similarity measure

| Pearson Similarity Measure | | | | |
|----|----|----|----|----|
|    | u2 | u3 | u4 | u5 |
| **u1** | 0.9487 | -0.7071 | -0.7921 | -1 |
| **u2** |  | -1 | -0.1886 | -1 |
| **u3** |  |  | 0 | -1 |
| **u4** |  |  |  | -0.5547 |

**Table 4** An example of the users similarity matrix, according to NHSM and Pearson similarity measures and Rank

|    | NHSM_Sim | | Pearson_Sim | |
|----|----|----|----|----|
|    | *Similarity* | *Rank* | *Similarity* | *Rank* |
| **u1-u2** | 0.0122 | 2 | 0.9487 | 1 |
| **u1-u3** | 0.0066 | 3 | -0.7071 | 2 |
| **u1-u4** | 0.0213 | 1 | -0.7921 | 3 |
| **u1-u5** | 0.0035 | 4 | -1 | 4 |

We discuss briefly the drawback of Pearson similarity measure that is used alone in most collaborative filtering recommender systems. First, according to Table 4, the similarity of u1 with u4 in the NHSM similarity measure is greater than u1 with u3, but this similarity is not accurate in the Pearson similarity measure. Using the NHSM similarity measure can overcome the drawback of the Pearson similarity measure (low similarity) regardless of similar ratings by two users. Second, if a high difference exists between the two user's ratings, the NHSM similarity will be very small. For example, the similarity between u2 and u5 is very low, it is about 0.0010. It is the least similarity in the user similarity matrix in Table 2. From Table 1, we can see that the rating vectors of these two users are (4, 4, ?, 2) and (1, ?, 3, ?) respectively. It has very low similarity indeed. The Pearson similarity measure cannot distinguish the similarity between users correctly. There are similarities -1 for some users in Table 3.

3.2 Prediction

In this section, The CUPCF calculates the amount of active user's preferences for her/his unrated items according to Eq. (11-12) based on each of the similarity measures.

$$NHSM\_Predict(u,p) = \mu_u + \frac{\sum_{j=1}^{m}(r_{v_j,p} - \mu_v).NHSM\_Sim(u,v_j)}{\sum_{j=1}^{m}|NHSM\_Sim(u,v_j)|} \quad (11)$$

$$Pearson\_Predict(u,p) = \mu_u + \frac{\sum_{j=1}^{m}(r_{v_j,p} - \mu_v).Pearson\_Sim(u,v_j)}{\sum_{j=1}^{m}|Pearson\_Sim(u,v_j)|} \quad (12)$$

Where $u$ is the active user and $p$ is an item that the CUPCF system is supposed to predict a rating for that. The CUPCF system calculates these ratings for all unobserved items by the active user and offers the $Top - N$ items. In Eq. (11), $\mu_u$ is the average ratings of active user and $m$ is the number of neighbor users with the active user. $NHSM\_Sim(u,v_j)$ and $Pearson\_Sim(u,v_j)$ are the similarity degree of the active user $u$ with the user $v_j$. $r_{v,p}$ is the rating of item $p$ by user $v$ and $\mu_v$ is the average ratings of user $v$.

Then, according to Eq. (13) as a new method, we combine the value of the predicted ratings. Given the data sparsity problem in rating matrix, it is possible that a certain similarity measure cannot predict a value, in this condition, most systems use the average of the active users' ratings for a target item. Therefore, by combining these two similarity measures as a new method, if the system is not able to predict the amount of user preferences by k-nearest users of a certain similarity measure, the system can use the other k-nearest users from another similarity measure to predict them. $CUP\_Predict(u,p)$ is a formula for combining user preferences is defined as follows:

$$CUP\_Predict(u,p) = \frac{NHSM\_Predict(u,p) + Pearson\_Predict(u,p)}{2} \quad (13)$$

Where $u$ is an active user and $p$ is an item. The $NHSM\_Predict(u,i)$ and the $Pearson\_Predict(u,p)$ are the predictions of item rating $p$ by Eq. (11) and Eq. (12). The output of the $NHSM\_Predict(u,i)$ and the $Pearson\_Predict(u,p)$ is between 1 to 5 and if each of the similarity measures are not able to predict the value of user preference, the $NHSM\_Predict(u,i)$ and the $Pearson\_Predict(u,p)$ return the average ratings of the active user. Finally, the $CUP\_Predict(u,p)$ uses similarity degree to predict the user preferences accurately.

3.3 Recommendation

In this section, a list of recommendations based on each similarity measure is created for active users. In this paper, we created a list of recommendations based on the $Top - N$ method. The values of N are 5, 10, 15, 20 and 30. Suppose that the list of recommendations is set to 5 (Top-5), the NHSM or Pearson similarity measure may not recommend 5 items to the active user. For example, from 5 items only recommended 3 items. So by combining these lists, you can increase the number of recommended items from 3 to 5.

## 4. Experimental Results

The performance of the proposed system was evaluated in the MovieLens dataset which consists of 943 users and 1682 items with 100,000 ratings for items [25, 26]. The ratings range in this dataset is from 1 to 5, which 5 being excellent and 1 being terrible.

To evaluate the system's performance, we tested on the dataset and used the 5-fold cross-validation algorithm, which provides 80% of the data for training and creating a proposed system model and 20% for system testing. The system evaluation based on the MAE, Accuracy, Precision and Recall metrics has been calculated according to Eq. (14-17) on the test data, which are shown in Table 5 of the related confusion matrix [27].

**Table 5** Confusion Matrix [27]

| Actual / Predicted | Negative | Positive |
|---|---|---|
| Negative | A | B |
| Positive | C | D |

$$MAE = \frac{\sum_{i=1}^{n}|\hat{r}_{u,p} - r_{u,p}|}{n} \tag{14}$$

$$Accuracy = \frac{Correct\ Recommendation}{Total\ Possible\ Recommedation} = \frac{A + D}{A + B + C + D} \tag{15}$$

$$Precision = \frac{Correctly\ Recommended\ Items}{Total\ Recommeded\ Items} = \frac{D}{B + D} \tag{16}$$

$$Recall = \frac{Correctly\ Recommended\ Items}{Total\ Useful\ Recommedations} = \frac{D}{C + D} \tag{17}$$

The proposed CUPCF system performs prediction and recommendation operations for each user separately in each fold. The number of $k$ for nearest neighbors in the calculation of prediction formula is set 300. The threshold value for calculating the Accuracy, Precision and Recall of $Top - Ns$ is set 3 and 4. The rating of 1 to 3 indicates an extreme dislike, and the rating of 3 to 5 indicates a strong affinity to the item. The threshold value(T) of 4 is the rating of 1 to 4 and 4 to 5. The N value of the highest items was selected $N = \{5, 10, 15, 20, 30\}$, respectively. CUPCF in terms of error rate is evaluated in 19 collaborative filtering algorithms and similarity measures such as: CF, CF-RA, CF-Diff, CF-Rank, CF-HW [15], CF-MW [15], Pearson, Ra-COS, RA-SRC, SRC, RA-CPC, CPC [18], Three-Segment, BCF, NHSM, PIP, COS [28], K-Means Leader and K-Means [29]. Fig. 3 shows the MAE of CUPCF in comparison with a number of state-of-the-art recommendation methods mentioned in references of [18, 28-29]. The percentage of improving of CUPCF than other models and techniques such as CF is 2.5%, CF-RA is 0.28%, CF-Diff is 0.8%, CF-Rank is 1.2%, CF-MW is 1.1%, CF-HW is 1.9%, Pearson is 1.5%, RA-COS is 1.4%, RA-SRC is 5.7%, SRC is 11.4%, RA-CPC is 1.1%, CPC is 2.3%, Three-Segment is 2.5%, BCF is 6.3%, NHSM is 12%, PIP is 15%, COS is 15.5%, K-Means Leader is 1.2% and K-Means is 3.2%. The other results for Accuracy, Precision and Recall of the CUPCF are shown in Tables 6 and 7. The maximum accuracy, precision and recall are in Top-5 items of Table 6, and the average of them are 0.91402, 0.91436 and 0.99744 respectively.

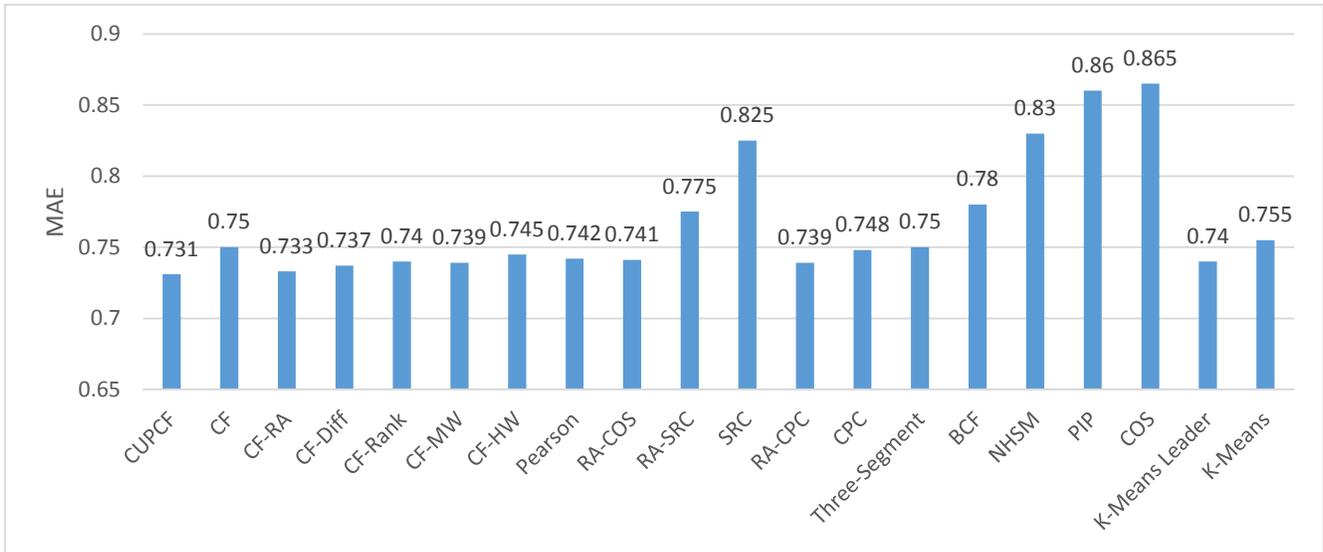

**Fig. 3** MAE of proposed system (CUPCF) with different collaborative filtering algorithms and techniques: collaborative filtering (CF), collaborative filtering with resource allocation (CF-RA), diffusion-based collaborative filtering (CF-Diff), rank-based collaborative filtering (CF-Rank), McLaughlin's weighting-based CF (CF-MW) [15], Herlocker's weighting-based CF (CF-HW) [15], Pearson Correlation (Pearson), Cosine similarity with RA (RA-COS), Spearman rank correlation with RA (RA-SRC), Spearman rank correlation (SRC), constrained Pearson correlation with RA (RA-CPC) and constrained Pearson correlation (CPC) [18], Three-Segment Similarity [28], BCF [28], NHSM [28], PIP [28], COS [28], K-Means Leader [29], K-Means [29].

**Table 6** System Evaluation Results, according to different Top-N with Threshold(T) = 3 in each fold

| Fold | Method Name | T | Evaluation Metric | Top 5 Items | Top 10 Items | Top 15 Items | Top 20 Items | Top 30 Items |
|---|---|---|---|---|---|---|---|---|
| 1 | CUPCF | 3 | Accuracy | 0.9171 | 0.9079 | 0.8994 | 0.8931 | 0.8826 |
|   |       |   | Precision | 0.9170 | 0.9092 | 0.9026 | 0.8980 | 0.8906 |
|   |       |   | Recall | 0.9982 | 0.9930 | 0.9875 | 0.9814 | 0.9721 |
|   |       |   | MAE | 0.7257 | 0.7257 | 0.7257 | 0.7257 | 0.7257 |
| 2 | CUPCF | 3 | Accuracy | 0.9160 | 0.9068 | 0.8977 | 0.8924 | 0.8822 |
|   |       |   | Precision | 0.9162 | 0.9083 | 0.9008 | 0.8967 | 0.8898 |
|   |       |   | Recall | 0.9975 | 0.9939 | 0.9887 | 0.9830 | 0.9733 |
|   |       |   | MAE | 0.7308 | 0.7308 | 0.7308 | 0.7308 | 0.7308 |
| 3 | CUPCF | 3 | Accuracy | 0.9140 | 0.9055 | 0.8989 | 0.8917 | 0.8813 |
|   |       |   | Precision | 0.9145 | 0.9080 | 0.9033 | 0.8985 | 0.8921 |
|   |       |   | Recall | 0.9975 | 0.9920 | 0.9860 | 0.9790 | 0.9685 |
|   |       |   | MAE | 0.7345 | 0.7345 | 0.7345 | 0.7345 | 0.7345 |
| 4 | CUPCF | 3 | Accuracy | 0.9117 | 0.9067 | 0.8985 | 0.8921 | 0.8817 |
|   |       |   | Precision | 0.9122 | 0.9084 | 0.9021 | 0.8980 | 0.8912 |
|   |       |   | Recall | 0.9966 | 0.9928 | 0.9872 | 0.9806 | 0.9706 |
|   |       |   | MAE | 0.7340 | 0.7340 | 0.7340 | 0.7340 | 0.7340 |
| 5 | CUPCF | 3 | Accuracy | 0.9113 | 0.9056 | 0.8975 | 0.8917 | 0.8817 |
|   |       |   | Precision | 0.9119 | 0.9065 | 0.9001 | 0.8955 | 0.8882 |
|   |       |   | Recall | 0.9974 | 0.9951 | 0.9899 | 0.9848 | 0.9762 |
|   |       |   | MAE | 0.7321 | 0.7321 | 0.7321 | 0.7321 | 0.7321 |

**Table 7** System Evaluation Results, according to different Top-N with Threshold(T) = 4 in each fold

| Fold | Method Name | T | Evaluation Metric | Top 5 Items | Top 10 Items | Top 15 Items | Top 20 Items | Top 30 Items |
|---|---|---|---|---|---|---|---|---|
| 1 | CUPCF | 4 | Accuracy | 0.7565 | 0.7385 | 0.7287 | 0.7224 | 0.7120 |
|   |       |   | Precision | 0.7599 | 0.7470 | 0.7386 | 0.7344 | 0.7257 |
|   |       |   | Recall | 0.9434 | 0.8963 | 0.8620 | 0.8339 | 0.7965 |
|   |       |   | MAE | 0.7257 | 0.7257 | 0.7257 | 0.7257 | 0.7257 |
| 2 | CUPCF | 4 | Accuracy | 0.7529 | 0.7375 | 0.7261 | 0.7205 | 0.7084 |
|   |       |   | Precision | 0.7611 | 0.7509 | 0.7428 | 0.7383 | 0.7294 |
|   |       |   | Recall | 0.9435 | 0.8990 | 0.8623 | 0.8360 | 0.7972 |
|   |       |   | MAE | 0.7308 | 0.7308 | 0.7308 | 0.7308 | 0.7308 |
| 3 | CUPCF | 4 | Accuracy | 0.7501 | 0.7343 | 0.7266 | 0.7202 | 0.7087 |
|   |       |   | Precision | 0.7558 | 0.7443 | 0.7390 | 0.7337 | 0.7249 |
|   |       |   | Recall | 0.9391 | 0.8937 | 0.8589 | 0.8338 | 0.7958 |
|   |       |   | MAE | 0.7345 | 0.7345 | 0.7345 | 0.7345 | 0.7345 |
| 4 | CUPCF | 4 | Accuracy | 0.7511 | 0.7338 | 0.7263 | 0.7192 | 0.7089 |
|   |       |   | Precision | 0.7552 | 0.7424 | 0.7363 | 0.7303 | 0.7228 |
|   |       |   | Recall | 0.9438 | 0.8991 | 0.8656 | 0.8370 | 0.7983 |
|   |       |   | MAE | 0.7340 | 0.7340 | 0.7340 | 0.7340 | 0.7340 |
| 5 | CUPCF | 4 | Accuracy | 0.7508 | 0.7324 | 0.7231 | 0.7193 | 0.7080 |
|   |       |   | Precision | 0.7505 | 0.7361 | 0.7278 | 0.7234 | 0.7138 |
|   |       |   | Recall | 0.9459 | 0.9050 | 0.8722 | 0.8477 | 0.8128 |
|   |       |   | MAE | 0.7321 | 0.7321 | 0.7321 | 0.7321 | 0.7321 |

Regarding the results of Table 6 and Table 7, we conclude that if we consider the number 3 for threshold (T) in calculating Accuracy, Precision and Recall, the CUPCF is able to recommend the user's favorite items with high accuracy, and the list of recommendations provided with a high percentage in accordance with the user's tastes. Although the results in Table 7 indicate the ability of the CUPCF to recommend the related items to users.

## 5. Conclusions

Recommender systems suggest items to users according to their implicit and explicit feedback information or users' preferences, such as ratings, reviews, and clicks. One of the most popular models used in these systems is collaborative filtering. This model suffers from the cold start and data sparsity problems. Therefore, we propose an effective recommender system (CUPCF) to solve them. This system uses the combination of the users' preferences in items prediction phase and recommends items which are close to active users' tastes. Therefore, CUPCF considers two similarity measures for finding the nearest neighbor users. Finally, the Top-N items are suggested by this system. The experimental results show that CUPCF error rate than a number of state-of-the-art recommendation systems and methods is improved and reduced. The percentage of improving of CUPCF than other models and techniques such as CF is 2.5%, CF-RA is 0.28%, CF-Diff is 0.8%, CF-Rank is 1.2%, CF-MW is 1.1%, CF-HW is 1.9%, Pearson is 1.5%, RA-COS is 1.4%, RA-SRC is 5.7%, SRC is 11.4%, RA-CPC is 1.1%, CPC is 2.3%, %, Three-Segment is 2.5%, BCF is 6.3%, NHSM is 12%, PIP is 15%, COS is 15.5%, K-Means Leader is 1.2% and K-Means is 3.2%. The maximum values accuracy, precision and recall are 0.91402, 0.91436 and 0.99744 respectively.

**Conflict of Interest:** The authors declare that they have no conflict of interest.

**Mostafa Khalaji** received the B.Sc. degree in computer engineering (Software) from Sadra Institute of Higher Education, Tehran, Iran, in 2015, and the M.Sc. degree in computer engineering (Artificial Intelligence) from K. N. Toosi University of Technology, Tehran, Iran, in 2017. He is a lecturer at Islamic Azad University, Shahr-e-Qods Branch and Sadra Institute of Higher Education. He is a member of IEEE. His current research interests include Recommender Systems, Machine Learning, Social Network Analysis, and Data mining. His ORCID ID is 0000-0002-5019-1824.

**Nilufar Mohammadnejad** is a B.Sc. student in computer engineering (Software) at Islamic Azad University, Shahr-e-Qods Branch, Tehran, Iran. Her current research interests include Recommender Systems.

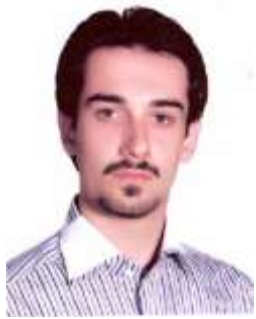

**Mostafa Khalaji**

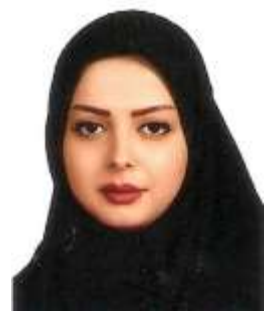

**Nilufar Mohammadnejad**